\documentclass[prl,twocolumn,tbtags,showpacs,superscriptaddress,floatfix]{revtex4}
\usepackage{amssymb}
\usepackage{graphicx,amsmath}
\usepackage{bm}
\usepackage{times}

\newcommand{\NLij}{i\kern -0.08em j}

\begin{document}

\title{Multiphoton excitations and inverse population in a system of two
flux qubits }

\begin{abstract}
We study the multiphoton spectroscopy of artificial solid-state four-level
quantum system. This system is formed by two coupled superconducting flux
qubits. When multiple driving frequency of the applied microwaves matches
the energy difference between any two levels, the transition to the upper
level is induced. We demonstrate two types of the multi-photon transitions:
direct transitions between two levels and ladder-type transitions via an
intermediate level. Our calculations show, that for the latter transitions,
in particular, the inverse population of the excited state with respect to
the ground one is realized. These processes can be useful for the control of
the level population for the multilevel scalable quantum systems.
\end{abstract}

\date{\today}
\author{E.~Il'ichev}
\affiliation{Institute of Photonic Technology, P.O. Box 100239, D-07702 Jena, Germany}
\author{S. N. Shevchenko}
\affiliation{B. Verkin Institute for Low Temperature Physics and Engineering, 47 Lenin
Ave., 61103 Kharkov, Ukraine}
\author{S. H. W. van der Ploeg}
\affiliation{Institute of Photonic Technology, P.O. Box 100239, D-07702 Jena, Germany}
\author{M.~Grajcar}
\affiliation{Institute of Photonic Technology, P.O. Box 100239, D-07702 Jena, Germany}
\affiliation{Department of Experimental Physics, Comenius University, SK-84248
Bratislava, Slovakia}
\author{E. A.~Temchenko}
\affiliation{B. Verkin Institute for Low Temperature Physics and Engineering, 47 Lenin
Ave., 61103 Kharkov, Ukraine}
\affiliation{Kharkov National University, 4 Svobody Sqr., 61077 Kharkov, Ukraine}
\author{A. N. Omelyanchouk}
\affiliation{B. Verkin Institute for Low Temperature Physics and Engineering, 47 Lenin
Ave., 61103 Kharkov, Ukraine}
\author{H.-G.~Meyer}
\affiliation{Institute of Photonic Technology, P.O. Box 100239, D-07702 Jena, Germany}
\maketitle


Manipulation and measurement of the state of scalable multilevel quantum
systems is one of the key issues for their implementation for quantum
information processing devices. One of the possible realizations of such
type of devices is based on superconducting Josephson junctions, where the
levels populations can be controlled by the external microwaves. When the
single or multiple driving frequency matches the energy difference between
energy levels, the transition to the upper level is induced. This provides
the tool for manipulation and characterization of the quantum system -
single/multiphoton spectroscopy.

Recently the spectroscopy of the energy levels in single Josephson-junction
qubits was demonstrated with multi-photon excitations by several groups (see
for example \cite{Oliver, Sillanpaa, Wilson07, Izmalkov08, Sun09}).
Moreover, similar multi-photon effects were studied in a single
superconducting circuit when it behaves as a multi-level system \cite{Yu,
Dutta08, Berns08}. On the other hand, the multi-level quantum system can be
realized in coupled qubits, where the one-photon driving was studied \cite%
{Pashkin03, Berkley03, 2qbs, Majer05, McDermott05, Fay07, Izmalkov08}. Such
two-qubit devices are of practical importance e.g. for building the
so-called universal set of gates \cite{You05, Wen-Shum07, Zag-Blais07} which
is essential part for realization of quantum computing \cite{Yamamoto,
Plantenberg07, Leek09}.

In this paper we report the observation of the multi-photon resonances in a
two-qubit system. The levels population in this effectively four-level
system can be controlled by external driving field. We demonstrate two types
of the resonant excitations: direct (from one level to another, when two
levels are relevant) and ladder-type (via an intermediate level, when three
levels are relevant). Moreover, our calculations show that the inverse
population takes place in our driven multi-level system. These results are
important for new, four-level quantum elements which are difficult to
realize in quantum optics. For example, they can generate extremely large
optical nonlinearities at microwave frequencies, with no associated
absorption \cite{Rebic}.

The investigated system consists of two coupled superconducting flux qubits.
A flux qubit is a superconducting ring with three Josephson junctions \cite%
{Mooij}. For controlling the state of the system, the microwave magnetic
flux is applied. For reading-out the state, the qubits are weakly coupled to
the resonant tank circuit.

The coupled flux qubits are described by the effective Hamiltonian in terms
of the Pauli matrices:
\begin{equation}
\widehat{H}=\sum\limits_{i=\mathrm{a,b}}\left( -\frac{\Delta _{i}}{2}%
\widehat{\sigma }_{x}^{(i)}-\frac{\epsilon _{i}(t)}{2}\widehat{\sigma }%
_{z}^{(i)}\right) +\frac{J}{2}\widehat{\sigma }_{z}^{(\mathrm{a})}\widehat{%
\sigma }_{z}^{(\mathrm{b})}.  \label{H}
\end{equation}%
The tunnelling amplitudes $\Delta _{i}$ and the coupling parameter $J$ both
are constants in this model. The biases
\begin{equation}
\epsilon _{i}(t)=2I_{\mathrm{p}}^{(i)}\Phi _{0}f^{(i)}(t)  \label{eps}
\end{equation}%
are controlled by the dimensionless magnetic fluxes $f^{(i)}(t)=\Phi
_{i}(t)/\Phi _{0}-1/2$ through $i$-th qubit:
\begin{equation}
f^{(i)}(t)=f_{i}+f_{\mathrm{ac}}\sin \omega t+f_{\mathrm{rf}}.  \label{fx}
\end{equation}%
Here $I_{\mathrm{p}}^{(i)}$ is the persistent current in the $i$-th qubit $%
f_{i}=\Phi _{\mathrm{dc}}^{(i)}/\Phi _{0}-1/2$, $f_{\mathrm{ac}}=\Phi _{%
\mathrm{ac}}/\Phi _{0}$, and $f_{\mathrm{rf}}$ stand for the flux introduced
by the tank coil. The parameters for the qubit system were obtained from the
ground-state measurements \cite{Izmalkov08}: $\Delta _{\mathrm{a}}=15.8$, $%
\Delta _{\mathrm{b}}=3.5$, $I_{\mathrm{p}}^{(\mathrm{a})}\Phi _{0}=375$, $I_{%
\mathrm{p}}^{(\mathrm{b})}\Phi _{0}=700$, $J=3.8$ in units of $h\cdot $GHz.

It was shown \cite{Greenberg02a, Smirnov03, QIMT}, that the response of the
measuring device, the tank circuit, is defined by the magnetic
susceptibility or the effective inductance of the qubit system. Namely, both
the phase shift and the voltage amplitude offset in the limit of small bias
current ($f_{\mathrm{rf}}\longrightarrow 0$) are proportional to $\lambda
=d\Phi _{\mathrm{tot}}/d\Phi _{\mathrm{x}}$, where $\Phi _{\mathrm{tot}%
}=\Sigma L_{i}I_{\mathrm{qb}}^{(i)}$ is the total magnetic flux and $%
d(...)/d\Phi _{\mathrm{x}}=\Sigma \partial (...)/\partial \Phi _{i}$ is the
symmetrized derivative. Here $L_{i}$\ and $I_{\mathrm{qb}}^{(i)}$ are the
geometrical inductance and the expectation value of the current in $i$-th
qubit. The value $\mathcal{L}_{i}^{-1}=dI_{\mathrm{qb}}^{(i)}/d\Phi _{%
\mathrm{x}}$ is the inverse effective inductance of the $i$-th qubit. Thus,
the tank voltage amplitude is related to%
\begin{equation}
\lambda =\left( \frac{\partial }{\partial \Phi _{\mathrm{a}}}+\frac{\partial
}{\partial \Phi _{\mathrm{b}}}\right) \left( L_{\mathrm{a}}I_{\mathrm{qb}}^{(%
\mathrm{a})}+L_{\mathrm{b}}I_{\mathrm{qb}}^{(\mathrm{b})}\right) .
\label{lambda}
\end{equation}

When a qubit is in the ground state, its current has a definite direction
and is defined by the energy derivative with respect to dc flux \cite%
{Greenberg02a}. Then the tank voltage amplitude (see Eq. (\ref{lambda})) as
a function of dc fluxes is defined by the second derivative of the energy,
i.e. by the curvature of the ground-state energy. Thus, one expects a dip in
the flux dependence of the tank response in the vicinity of the energy
avoided crossing \cite{Greenberg02a, Grajcar05, th}. In the driven qubits
the current has the probabilistic character. As the consequence, equation (%
\ref{lambda}) includes not only the terms with energy derivative, but also
the terms with the probability derivative \cite{Sillanpaa, QIMT}. When the
system is driven, its upper level occupation probability is resonantly
increased. The respective derivative is displayed as the alteration of peak
and dip centered around the resonance \cite{Grajcar07, th}. Correspondingly,
if the tank voltage is plotted as a function of two parameters (two partial
dc flux biases), the resonances appear as the alteration of ridges and
troughs.

\begin{widetext}

\begin{figure}[h]
\includegraphics[width=16cm]{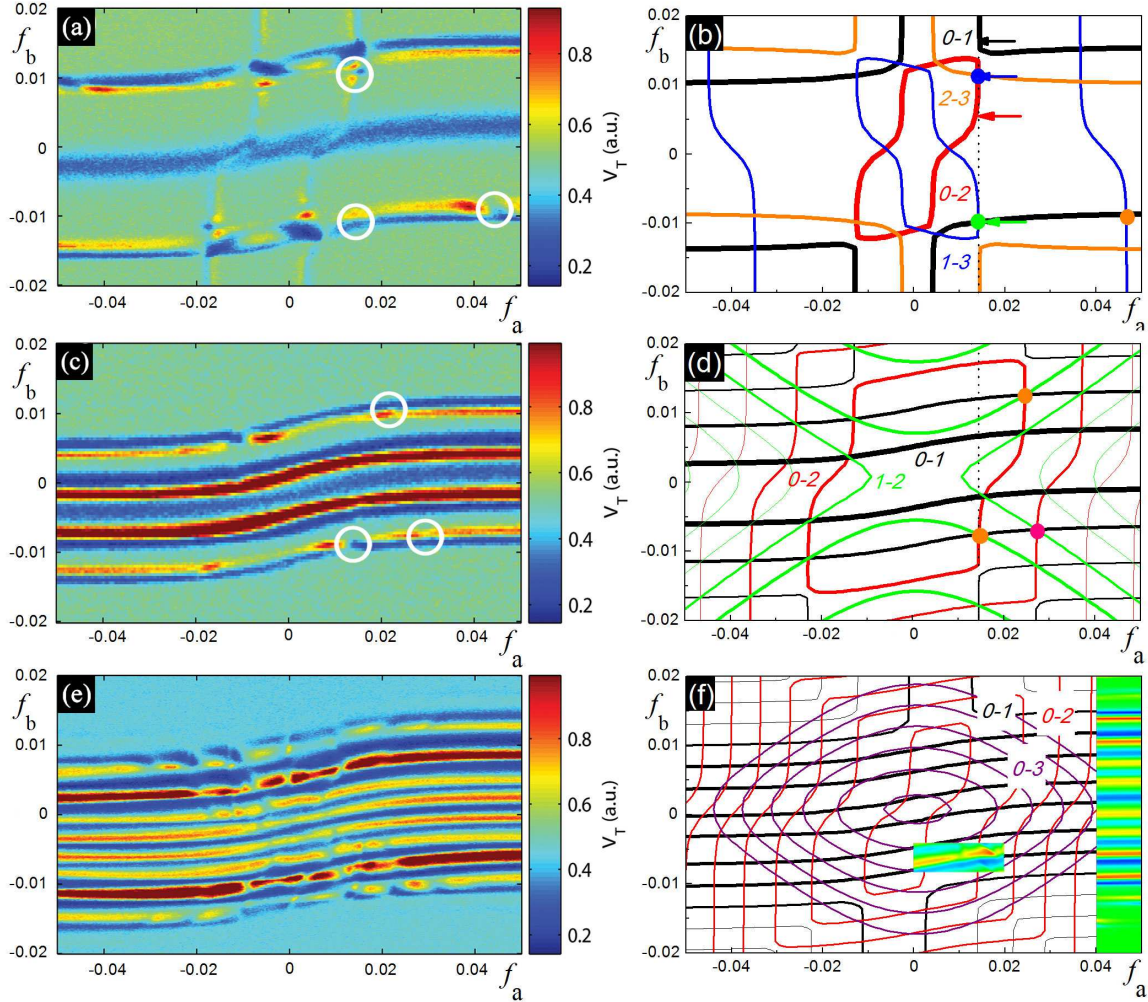}
\caption{(Color online). Multi-photon resonances in the system of
coupled flux qubits. The tank voltage amplitude (a, c, e) and the
energy contour lines (b, d, f) versus the partial bias fluxes
$f_{\mathrm{a}}$ and $f_{\mathrm{b}}$ for
different driving frequencies (from top to bottom): $\protect\omega /2%
\protect\pi =17.6$, $7.0$, $4.1$ GHz. Numbers $k-j$ next to the
lines mean that the line relates to the energy difference
$E_{j}-E_{k}$. The color inset in (f) is plotted for $\lambda$,
Eq.~(\ref{lambda}), by solving the Bloch-Redfield equation for the
reduced density matrix.} \label{Fig1}
\end{figure}

\end{widetext}

In the left panel in Fig.~\ref{Fig1} we present the experimental results
(the voltage amplitude of the tank as a function of qubit biases) for the
system of two coupled flux qubits. The driving frequency $\omega /2\pi $\ is
$17.6$ for (a), $7.0$ GHz for (c), and $4.1$ for (e). The system can be
resonantly excited from the level $k$ to the level $j$ when the energy of $n$
photons matches them:
\begin{equation}
\Delta E_{kj}(f_{\mathrm{a}},f_{\mathrm{b}})\equiv E_{j}-E_{k}=n\cdot \hbar
\omega .  \label{E_ij}
\end{equation}%
Then along the contour lines defined with this relation the resonant
structure appears. (Besides, the trough due to the ground state curvature is
visible at the center, around $f_{\mathrm{b}}$ close to $0$.) The resonances
are visualized with the ridge-trough line. However, the ridge-trough line is
disturbed with increasing or decreasing the signal; some of these changes
are shown with white circles. This means changing the effective Josephson
inductance in these points. We argue that this happens because of the
ladder-type multi-photon excitations to higher levels (see also Ref. \cite%
{Fink08}).%

\begin{figure}[h]
\includegraphics[width=8cm]{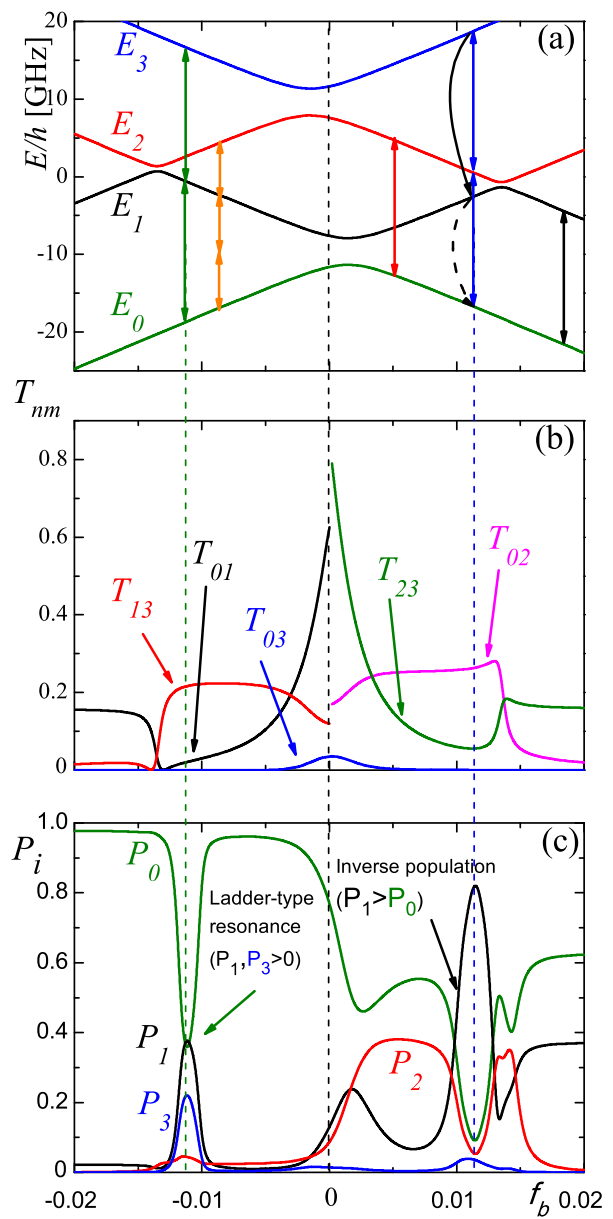}
\caption{(Color online). Ladder-type transitions and the inverse population.
(a) Energy levels of two coupled qubits at $f_{\mathrm{a}}=0.015$ versus $f_{%
\mathrm{b}}$ (\textit{i.e.} along the dotted line in Fig.~\protect\ref{Fig1}
(b,d)). (b) Transition matrix elements $T_{nm}$ between the eigenstates $%
\left\vert E_{m}\right\rangle $ and $\left\vert E_{n}\right\rangle $. (c)
The energy levels occupation probabilities $P_{i}$ under driving with
frequency $\protect\omega /2\protect\pi =17.6$ GHz.}
\label{Fig2}
\end{figure}

To understand the experimental results, we plot the energy contour lines in
Fig.~\ref{Fig1}(b,d,f) for three frequencies $\omega $, by making use of
relation (\ref{E_ij}). Consider first Fig.~\ref{Fig1}(b). The black and red
lines show the positions of the expected resonant excitations from the
ground state to the first and to the second excited states respectively. The
blue and orange lines are the contour lines for the possible excitations
from the first and from the second excited state to the third excited state.
The \textit{one-photon resonances} along the black and red curves are
clearly visible in Fig.~\ref{Fig1}(a); this was used for the spectroscopical
study of the system \cite{Izmalkov08}. For better understanding we calculate
the energy levels, by diagonalizing Hamiltonian (\ref{H}), and plot them at
fixed value of the bias flux through qubit $a$, $f_{\mathrm{a}}$, as a
function of the bias flux through qubit $b$, $f_{\mathrm{b}}$, in Fig.~\ref%
{Fig2}(a); the arrows of the length $\omega /2\pi =17.6$ GHz and $7.0$ GHz
(orange) are introduced to match the energy levels. The black and red arrows
in both Fig.~\ref{Fig2}(a) and Fig.~\ref{Fig1}(b) show the position of
one-photon transitions to the first and the second excited levels. The
double green and blue arrows in Fig.~\ref{Fig2} show the position of the
two-photon processes, where the excitation by the first photon creates the
population of the first and the second levels and the second photon excites
the system to the upper level. We emphasize that \textit{these} \textit{%
two-photon excitations happen via intermediate levels}. The position of
these expected resonances is shown in Fig.~\ref{Fig1}(b) with the blue and
green thick points. Indeed, there is the change of the signal in Fig.~\ref%
{Fig1}(a) in these points. (The two-photon resonance $0-2-3$, shown with the
blue point, was shown with the numerical simulation in Ref. \cite{th}.)
Moreover, the orange point in Fig.~\ref{Fig1}(b) stand for the \textit{%
ladder-type three-photon excitation}, $0-1--3$ (with one photon to the first
excited level and then with two photons to the upper level), which is also
visible in Fig.~\ref{Fig1}(a).

In Fig.~\ref{Fig1}(c) we can see the ridge-trough resonances for the driving
frequency $\omega /2\pi =7$ GHz. Comparing with the contour lines in Fig.~%
\ref{Fig1}(d) we easily notice that the resonances are one- and two-photon
resonant excitations to the first excited level. Note that the two-photon
resonant excitation is direct and happen without any intermediate level in
contrast to the above described resonances. The higher level excitations via
the first excited state appear due to \textit{three- and four-photon
excitations}, as shown with orange ($0--1-2$) and pink ($0--1--2$) dots in
Fig.~\ref{Fig1}(d). These resonances are visible in Fig.~\ref{Fig1}(c).

In Fig.~\ref{Fig1}(e) the response of the two-qubit system is shown for $%
\omega /2\pi =4.1$ GHz. The lines along the horizontal axis are due to
\textit{direct 1-, 2-, 3-, and 4-photon excitations to the first excited
state}; cf. black lines in Fig.~\ref{Fig1}(f). Numerous \textit{upper level
excitations via the first excited level }appear as the amplification and
lowering the signal along these lines. For illustration in Fig.~\ref{Fig1}%
(f) beside transitions to the first excited level (black line) we plot the
red and violet lines which match the ground state and the second and third
excited states.

The transition probability from the state $\left\vert E_{m}\right\rangle $
into the state $\left\vert E_{n}\right\rangle $ is defined by the absolute
value of the matrix element of the perturbation. From Eqs.~(\ref{H}-\ref{fx}%
) these are given by:%
\begin{eqnarray}
T_{nm} &=&\left\vert \left\langle E_{n}\right\vert \widehat{v}\left\vert
E_{m}\right\rangle \right\vert ^{2}, \\
\widehat{v} &=&\frac{1}{I_{\mathrm{p}}^{(\mathrm{b})}}\left( I_{\mathrm{p}%
}^{(\mathrm{a})}\widehat{\sigma }_{z}^{(\mathrm{a})}+I_{\mathrm{p}}^{(%
\mathrm{b})}\widehat{\sigma }_{z}^{(\mathrm{b})}\right) ,  \notag
\end{eqnarray}%
where we have divided the perturbation by the factor $I_{\mathrm{p}}^{(%
\mathrm{b})}\Phi _{0}\cdot f_{\mathrm{ac}}$. We plot the transition matrix
elements in Fig.~\ref{Fig2}(b) to describe the ladder-type excitations shown
in Fig.~\ref{Fig1}(b) with the green and blue points. The position of the
respective resonances is shown in Fig.~\ref{Fig2}(a) and (b) respectively
with green and blue arrows and dotted lines. In the left half of Fig.~\ref%
{Fig2}(b) we plot the transition elements around the transition between the
three levels $E_{0}$, $E_{1}$, and $E_{3}$. In the right half we plot the
transition elements related to three levels $E_{0}$, $E_{2}$, and $E_{3}$.
In the latter case the transition element between the higher two levels ($%
E_{2}$ and $E_{3}$) is smaller than between the lower two levels ($E_{0}$
and $E_{2}$): $T_{23}\ll T_{02}$. In contrast, in the former case the
transition element between the higher two levels ($E_{1}$ and $E_{3}$) is
significantly larger than between the lower two levels ($E_{0}$ and $E_{1}$%
): $T_{13}\gg T_{01}$. Important to note that in both cases the transition
element from the ground state to the highest excited level, $T_{03}$, is
very small. This means that the probability of the direct excitation to the
highest level is very small -- the transition is induced exceptionally due
to ladder-type mechanism.

And finally we calculate the energy level occupation probabilities under
driving. This is done by means of the numerical solution of the
Bloch-Redfield equation for the qubit system density matrix $\widehat{\rho }$
\cite{Storcz03}. The impact of the dissipative environment on the qubit
system is described by the Redfield relaxation tensor. Conveniently the
environment is modelled as the harmonic oscillator common bath with the
ohmic spectral densities \cite{Storcz03}. Then the coupling of the qubit
system to the environment is characterized with only one phenomenological
parameter $\alpha $, which describes the strength of the dissipative
effects. Important to note that this model accurately describes the
relaxation between different levels of the systems (in contrast to the often
used approach with equal relaxation rates for all energy levels), since
respective relaxation rates are essentially different. Then the fitting of
the experimental graphs is done with Eq.~(\ref{lambda}), where the
expectation value of the current in $i$-th qubit is calculated with the
reduced density matrix: $I_{\mathrm{qb}}^{(i)}=-I_{\mathrm{p}}^{(i)}Sp(%
\widehat{\rho }\widehat{\sigma }_{z}^{(i)})$. The result of the calculation
is presented as the inset in Fig.~\ref{Fig1}(f). Such fitting gives us the
parameter for the strength of dissipation $\alpha =0.1$ and the driving
amplitudes $f_{\mathrm{ac}}\times 10^{3}=4,$ $6,$ and $8$ for Fig.~\ref{Fig1}%
(a), (c), and (e) respectively.

The numerically calculated energy level occupation probabilities are plotted
in Fig.~\ref{Fig2}(c) for $\omega /2\pi =17.6$ GHz, $f_{\mathrm{ac}}=4\times
10^{-3}$, $f_{\mathrm{a}}=15\times 10^{-3}$ versus $f_{\mathrm{b}}$, that is
along the dotted line in Fig.~\ref{Fig1}(b). This graph demonstrates two
interesting phenomena, similar to those which exhibit atoms in the laser
field \cite{Vitanov01}. First, the \textit{ladder-type resonant excitation}
takes place to the left, where the upper level occupation probability $P_{3}$
is of the same order as the intermediate level occupation probability $P_{1}$%
. This corresponds to the green point in Fig.~\ref{Fig1}(b). Second, the
\textit{inverse population }appears to the right (corresponds to the blue
point in Fig.~\ref{Fig1}(b)). This means that the upper level occupation
probability $P_{1}$ is larger than the ground state one $P_{0}$ \cite%
{Astafiev07, You07, Berns08, Sun09}. In our case the inverse population at
the first excited level is accumulated after relaxation from the third
excited level, since relaxation from the third level to the first one (shown
with solid curved arrow in Fig.~\ref{Fig2}(a)) is larger than both
relaxation from the third to second and from the first to the ground state
(shown with dashed curved arrow in Fig.~\ref{Fig2}(a)).

In conclusion, the multi-photon resonances in the four-level (two flux
qubit) system were observed. The multi-photon resonances were of two kinds:
direct and via intermediate level (when three levels are relevant for the
process). Our calculations show that in the latter case the stationary state
of the system with different relaxation rates exhibits the inverse
population. Such effects are relevant for both multi-photon spectroscopy of
the system and for the study of new effects in artificial multi-level
structures.

This work was supported by the EU through the EuroSQIP project, by the DFG
project IL 150/6-1 and by Fundamental Researches State Fund grant F28.2/019.
E.I. acknowledges the financial support from Federal Agency on Science and
Innovations of Russian Federation under contract No. 02.740.11.5067 and the
financial support from Russian Foundation for Basic Research, Grant
RFBR-FRSFU No. 09-02-90419. M.G. was partially supported by the Slovak
Scientific Grant Agency Grant No. 1/0096/08, the Slovak Research and
Development Agency under the contract No. APVV-0432-07 and No.
VCCE-005881-07, ERDF OP R\&D, Project CE QUTE ITMS 262401022, and via CE SAS
QUTE.


\begin{thebibliography}{99}
\bibitem{Oliver} W.D. Oliver, Ya. Yu, J.C. Lee, K.K. Berggren, L.S. Levitov,
and T.P. Orlando, Science \textbf{310}, 1653 (2005).

\bibitem{Sillanpaa} M. Sillanp\"{a}\"{a}, T. Lehtinen, A. Paila, Yu.
Makhlin, and P. Hakonen, Phys. Rev. Lett. \textbf{96}, 187002 (2006).

\bibitem{Wilson07} C.M. Wilson, T. Duty, F. Persson, M. Sandberg, G.
Johansson, and P. Delsing, Phys. Rev. Lett. \textbf{98}, 257003 (2007).

\bibitem{Izmalkov08} A. Izmalkov, S.H.W. van der Ploeg, S.N. Shevchenko, M.
Grajcar, E. Il'ichev, U. H\"{u}bner, A.N. Omelyanchouk, and H.-G. Meyer,
Phys. Rev. Lett. \textbf{101}, 017003 (2008).

\bibitem{Sun09} G. Sun, X. Wen, Y. Wang, Sh. Cong, J. Chen, L. Kang, W. Xu,
Y. Yu, S. Han, and P. Wu, Appl. Phys. Lett. \textbf{94}, 102502 (2009).

\bibitem{Yu} Y. Yu, W.D. Oliver, J.C. Lee, K.K. Berggren, L.S. Levitov, and
T.P. Orlando, arXiv:cond-mat/0508587.

\bibitem{Dutta08} S.K. Dutta, F.W. Strauch, R.M. Lewis, K. Mitra, H. Paik,
T.A. Palomaki, E. Tiesinga, J.R. Anderson, A.J. Dragt, C.J. Lobb, and F. C.
Wellstood, Phys. Rev. B \textbf{78}, 104510 (2008).

\bibitem{Berns08} D.M. Berns, M.S. Rudner, S.O. Valenzuela, K.K. Berggren,
W.D. Oliver, L.S. Levitov, and T.P. Orlando, Nature \textbf{455}, 51 (2008).

\bibitem{Pashkin03} Yu.A. Pashkin, T. Yamamoto, O. Astafiev, Y. Nakamura, D.
V. Averin, and J. S. Tsai, Nature \textbf{421}, 823 (2003).

\bibitem{Berkley03} A.J. Berkley, H. Xu, R.C. Ramos, M.A. Gubrud, F. W.
Strauch, P. R. Johnson, J. R. Anderson, A. J. Dragt, C.J. Lobb, and F.C.
Wellstood, Science \textbf{300}, 1548 (2003).

\bibitem{2qbs} A. Izmalkov, M. Grajcar, E. Il'ichev, Th. Wagner, H.-G.
Meyer, A.Yu. Smirnov, M.H.S. Amin, A. Maassen van den Brink, and A.M.
Zagoskin, Phys. Rev. Lett. \textbf{93}, 037003 (2004).

\bibitem{Majer05} J.B. Majer, F.G. Paauw, A.C.J. ter Haar, C.J.P.M. Harmans,
and J.E. Mooij, Phys. Rev. Lett. \textbf{94}, 090501 (2005).

\bibitem{McDermott05} R. McDermott, R.W. Simmonds, M. Steffen, K.B. Cooper,
K. Cicak, K. D. Osborn, S. Oh, D.P. Pappas, and J. M. Martinis, Science
\textbf{307}, 1299 (2005).

\bibitem{Fay07} A. Fay, E. Hoskinson, F. Lecocq, L.P. Levy, F.W.J. Hekking,
W. Guichard, and O. Buisson, Phys. Rev. Lett. \textbf{100}, 187003 (2008).

\bibitem{You05} J.Q. You and F. Nori, Physics Today \textbf{58}(11), 42
(2005).

\bibitem{Wen-Shum07} G. Wendin and V.S. Shumeiko, Low Temp. Phys. \textbf{33}%
, 724 (2007).

\bibitem{Zag-Blais07} A. Zagoskin and A. Blais, Phys. Canada \textbf{63},
215 (2007).

\bibitem{Yamamoto} T. Yamamoto, Yu. A. Pashkin, O. Astafiev, Y. Nakamura,
and J. S. Tsai, Nature \textbf{425}, 941 (2003).

\bibitem{Plantenberg07} J.H. Plantenberg, P.C. de Groot, C.J.P.M. Harmans,
and J.E. Mooij, Nature \textbf{447}, 836 (2007).

\bibitem{Leek09} P.J. Leek, S. Filipp, P. Maurer, M. Baur, R. Bianchetti,
J.M. Fink, M. G\"{o}ppl, L. Steffen, and A. Wallraff, Phys. Rev. B \textbf{79%
}, 180511(R) (2009).

\bibitem{Rebic} S. Rebi\'{c}, J. Twamley, and G. J. Milburn, Phys. Rev.
Lett. \textbf{103}, 150503 (2009).

\bibitem{Mooij} J.E. Mooij, T.P. Orlando, L. Levitov, L. Tian, C.H. van der
Wal, and S. Lloyd, Science \textbf{285}, 1036 (1999).

\bibitem{Greenberg02a} Ya.S. Greenberg, A. Izmalkov, M. Grajcar, E.
Il'ichev, W. Krech, H.-G. Meyer, M.H.S. Amin, and A. Maassen van den Brink,
Phys. Rev. B \textbf{66}, 214525 (2002).

\bibitem{Smirnov03} A.Yu. Smirnov, Phys. Rev. B \textbf{68}, 134514 (2003).

\bibitem{QIMT} S.N. Shevchenko, Eur. Phys. J. B \textbf{61}, 187 (2008).

\bibitem{Grajcar05} M. Grajcar, A. Izmalkov, S.H.W. van der Ploeg, S.
Linzen, E. Il'ichev, Th. Wagner, U. H\"{u}bner, H.-G. Meyer, A. Maassen van
den Brink, S. Uchaikin, and A.M. Zagoskin, Phys. Rev. B \textbf{72},
020503(R) (2005).

\bibitem{th} S.N.~Shevchenko, S.H.W.~van~der~Ploeg, M.~Grajcar, E.~Il'ichev,
A.N.~Omelyanchouk, and H.-G.~Meyer, Phys. Rev. B \textbf{78}, 174527 (2008).

\bibitem{Grajcar07} M. Grajcar, S.H.W. van der Ploeg, A. Izmalkov, E.
Il'ichev, H.-G. Meyer, A. Fedorov, A. Shnirman, and G. Sch\"{o}n, Nature
Phys. \textbf{4}, 612 (2008).

\bibitem{Fink08} J.M Fink, M. G\"{o}ppl, M. Baur, R. Bianchetti, P.J. Leek,
A. Blais, and A. Wallraff, Nature \textbf{454}, 315 (2008).

\bibitem{Storcz03} M.J. Storcz and F.K. Wilhelm, Phys. Rev. A \textbf{67},
042319 (2003).

\bibitem{Vitanov01} N.V. Vitanov, T. Halfmann, B.W. Shore, and K. Bergmann,
Annu. Rev. Phys. Chem. \textbf{52}, 763 (2001).

\bibitem{Astafiev07} O. Astafiev, K. Inomata, A.O. Niskanen, T. Yamamoto,
Yu.A. Pashkin, Y. Nakamura, and J.S. Tsai, Nature \textbf{449}, 588 (2007).

\bibitem{You07} J.Q. You, Yu-xi Liu, C.P. Sun, and F. Nori, Phys. Rev. B
\textbf{75}, 104516 (2007).
\end{thebibliography}
\end{document}